# Reduction of Surgical Risk Through the Evaluation of Medical Imaging Diagnostics

Marco A. V. M. Grinet[a,*], Nuno M. Garcia[a], Ana I. R. Gouveia[b], José A. F. Moutinho[b], Abel J. P. Gomes[a]

[a] Department of Computer Science & Engineering, University of Beira Interior, Covilhã, Portugal
[b] Faculty of Health Sciences, University of Beira Interior, Covilhã, Portugal

ABSTRACT

*Objective*: Computer-aided diagnosis (CAD) of Breast Cancer (BRCA) images has been an active area of research in recent years. One of the main goals of this research area is to develop reliable automatic methods for detecting and diagnosing different types of BRCA from diagnostic images. In this paper, we present a review of the state-of-the-art CAD methods applied to magnetic resonance imaging (MRI) and mammography images of BRCA patients. The review aims to: (i) provide an extensive introduction to different radiomic features extracted from BRCA images through texture and statistical analysis and (ii) to categorize and group deep learning frameworks and data structures capable of using metadata to aggregate relevant information that can assist oncologists and radiologists with a CAD.

*Methods and material*: We divide the existing literature according to the imaging modality (mammography and MRI) and into radiomic methods, machine learning methods, or combination of both. We also emphasize the difference between each imaging modality and method's strengths and weaknesses and analyze their performance in detecting BRCA through a quantitative comparison.

*Results*: We compare the results of various approaches for implementing computer-aided diagnosis systems for the detection of BRCA. Each approach's standard workflow components are reviewed and summary tables provided. We present an extensive literature review of radiomics feature extraction techniques and machine learning methods applied in BRCA diagnosis and detection, focusing on data-preparation, data-structures, pre-processing and post-processing strategies available in the literature.

*Conclusions*: There is a growing interest on radiomic feature extraction and machine learning methods for BRCA detection through histopathological images, MRI and mammography images. However, there is not yet a single CAD method able to combine distinct data types to provide the best diagnostic results. Employing data-fusion techniques to medical images and patient data could lead to improved detection and classification results.

# Table of Contents

# 1. Introduction

Breast cancer is a major cause of death for women around the world. In 2019, it an estimated 268,600 women will be diagnosed with invasive breast cancer (BRCA) and 62,930 women will be diagnosed with non-invasive BRCA, in the United States alone [1]. Early diagnosis of breast carcinomas can greatly improve the long-term survival rates. Medical imaging has been a powerful tool used for early cancer detection, as well as for monitoring the patient during and after treatment or surgery. The three main imaging modalities used for cancer detection and classification are (i) histopathological images, (ii) mammography images, and (iii) magnetic resonance images (MRI). The trend of scientific publications can be observed in Figure 1, in which the number of papers using the methods discussed in this review has increased yearly since 2015. To standardize cancer diagnosis, the American College of Radiology (ACR) has implemented the Breast Imaging Report and Data system (BIRADS) for breast cancer image diagnostics. Similarly, the Union for International Cancer Control (UICC) has implemented the Classification of Malignant Tumors (TNM) for histopathological images. Currently, breast cancer screening is the best way to diagnose the disease in early stages. However, manual classification within these systems can be subject to bias and human error.

Breast cancer screening, more specifically through mammography, is a proven method to reduce mortality of breast cancer between 10% and 30% [2, 3]. The sensitivity of mammography examinations is reduced in younger, pre-menopausal woman, as well as women with dense breast tissue [4, 5]. In some cases, when the patient presents palpable breast masses, mammography has a false-negative rate between 8% and 16% [6, 7]. Some studies show an increased rate of false-negative results for women aged 40 to 79 years with first-degree relatives with breast cancer. Women with previous benign breast biopsy results aged 50 to 89 years and women with lower BMI (30kg/m$^2$) aged 50 to 59 years showed higher rates of false-negative [8]. MRI examination mitigates the problems of mammography, with increased detection sensitivity, and is often used in conjunction with mammography in breast cancer screenings. However, MRI screening is only applicable to women with elevated risks of breast cancer, such as genetic predisposition or dense breast tissue, due to its high cost.

According to Bray et al. [9], there will be an estimated increase of 75% in the number of cancer cases by 2030. Early detection of cancer is a key factor in reducing the mortality rate [10]. Currently, imaging modalities are used to detect cancer, however biopsy is still the most applied method used by clinicians to diagnose cancer and determine the type of tumor [2]. Current literature shows that a high number of BRCA screenings are false-negatives [6-8].

In this report we present a review of research done in the field of CAD methods for the detection and diagnosis of BRCA images with emphasis on histopathological images, MRI and mammography images, as well as present our results in the development of BRCA segmentation algorithms, as presented in the original objectives during the duration of the Santander grant.

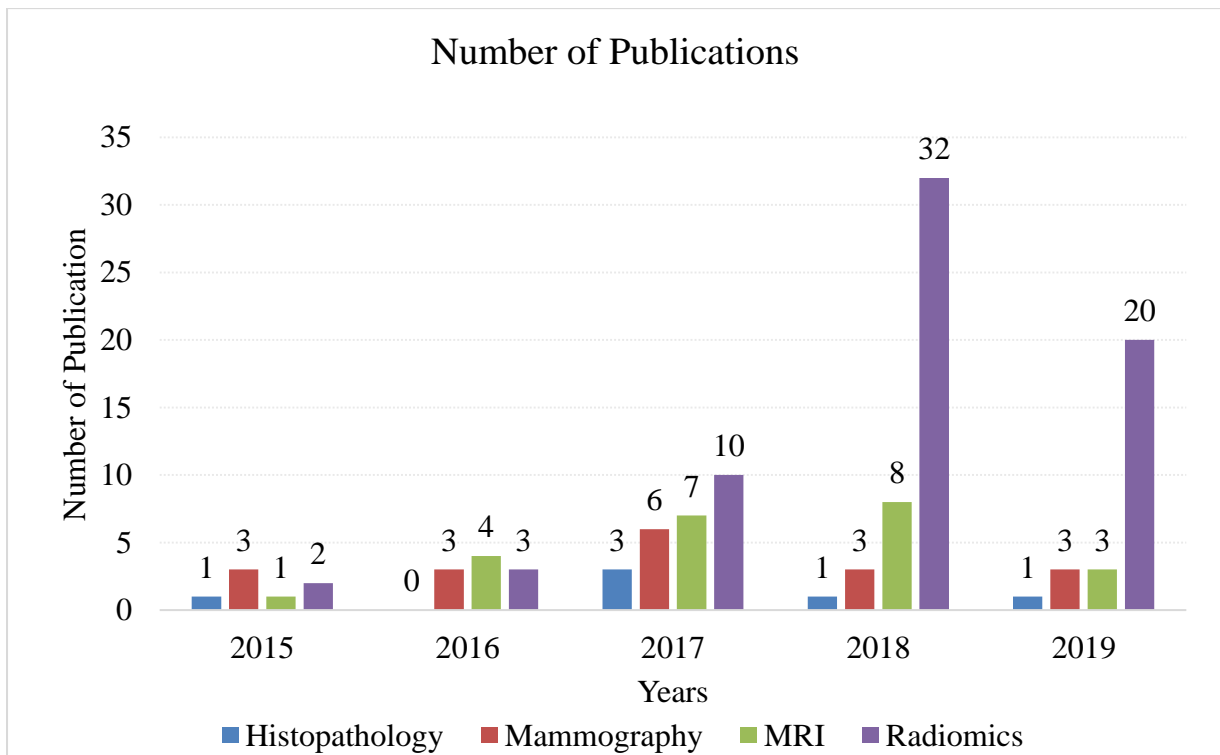

**Figure 1.** Number of publications per year in Pubmed containing "machine learning", "breast cancer" and one of the four modality keywords (histopathology, MRI, mammography, radiomics) from 2015 to 2019. (Queried: July 10th, 2019).

## 2. Breast Cancer Imaging

For over 30 years, medical imaging techniques have been used to assist doctors in the accurate detection and classification of tumors. More recently, several different cancer imaging modalities have been employed to assist oncologists BRCA diagnosis. These state-of-the-art methods and techniques used in BRCA diagnosis apply advanced image processing algorithms to extract useful information from the MRI and mammography images. Some of the more prominent methods in the literature include, Wavelet Transform (Single, Stationary, Discrete) [11], image fuzzyfication/defuzzyfication [12, 13], and image fusion techniques. This growing variety of cancer imaging methods [14 – 16] has led to the development of novel indicators for breast cancer risk factors and early diagnostics [17, 18]. Mammography is the most commonly used imaging modality for breast cancer screening and it has proven to be a useful tool in the early detection of the disease [19]. However, breast density has been shown to pose a challenge in mammograms, making cancer more difficult to detect in women with dense breasts [20]. In many cases, a secondary diagnostic method is employed to confirm the mammography results, such as MRI, histopathological image analysis, and biopsies [21 – 23].

MRI has been commonly used to detect and characterize breast lesions. In comparison to other imaging methods, MRI possesses a superior detection sensitivity for tumors, as well as metastasis [24 - 27], and has been shown to outperform ultrasound and mammography, when combined [20]. Recently, the literature shows that MRI can be an effective modality with good sensitivity for breast cancer [28 - 30]. However, tumor segmentation, in any imaging modality, is a process that is still widely performed in a manual or semi-automatic manner, and therefore is prone to human error [31 – 32].

Medical imaging techniques provide an enormous amount of patient and disease information. For breast cancer imaging, specially, multiple modalities are commonly used to reinforce a diagnosis, which leads to an even larger amount of data. The field of radiomics has been recently introduced as a solution to extract and compile the high-dimensional quantitative data obtained from imaging modalities into a more robust and descriptive analysis of tumors [28, 29]. Some examples of the application of radiomics is in patient diagnosis [15, 16], as well as patient prognosis [17 - 19]. Therefore, radiomics can be a powerful asset in breast cancer staging, as well as assist in surgical planning and treatment strategies. In the past few years, radiomics has shown to be an efficient method for high-throughput feature extraction from diagnostic images [30 - 32]. Machine learning methods can be employed to process these extracted features, and different machine learning architectures can be applied to identify and classify cancer, as well as predict cancer progression in patients. However, to extract relevant features from imaging data, it is necessary to have a complete understanding of the disease to avoid redundancy in the features and reduce dimensionality. Previous work has been done to evaluate and compare distinct feature selection methods and their impact on the classification of cancer patients [32].

*2.1.* Histopathological Images

Histopathology is an imaging modality that provides a visual examination of microscopic cellular tissue. This is an invasive diagnostic modality and is commonly used after less invasive imaging diagnostics to confirm a diagnosis, and requires a biopsy. In the biopsy, a small tissue sample is collected from the patient and placed into slides to perform the diagnosis. This imaging modality assists pathologists in BRCA grading, revealing important characteristics of the type and stage of the disease. The Nottingham Grading System (NGS) is currently the most commonly used method used by laboratories and clinicians for BRCA grading. More recently, with the growing implementation of digital slide scanners, imaging processing as well as machine learning algorithms can be employed to assist the grading of BRCA. Some notable datasets for histopathological BRCA images are the Breast Cancer Histopathological Database (BreakHis) and the Breast Cancer Histopathological Annotation and Diagnosis Dataset (BreCaHAD). Figure 2 shows an excerpt from the BreakHis dataset.

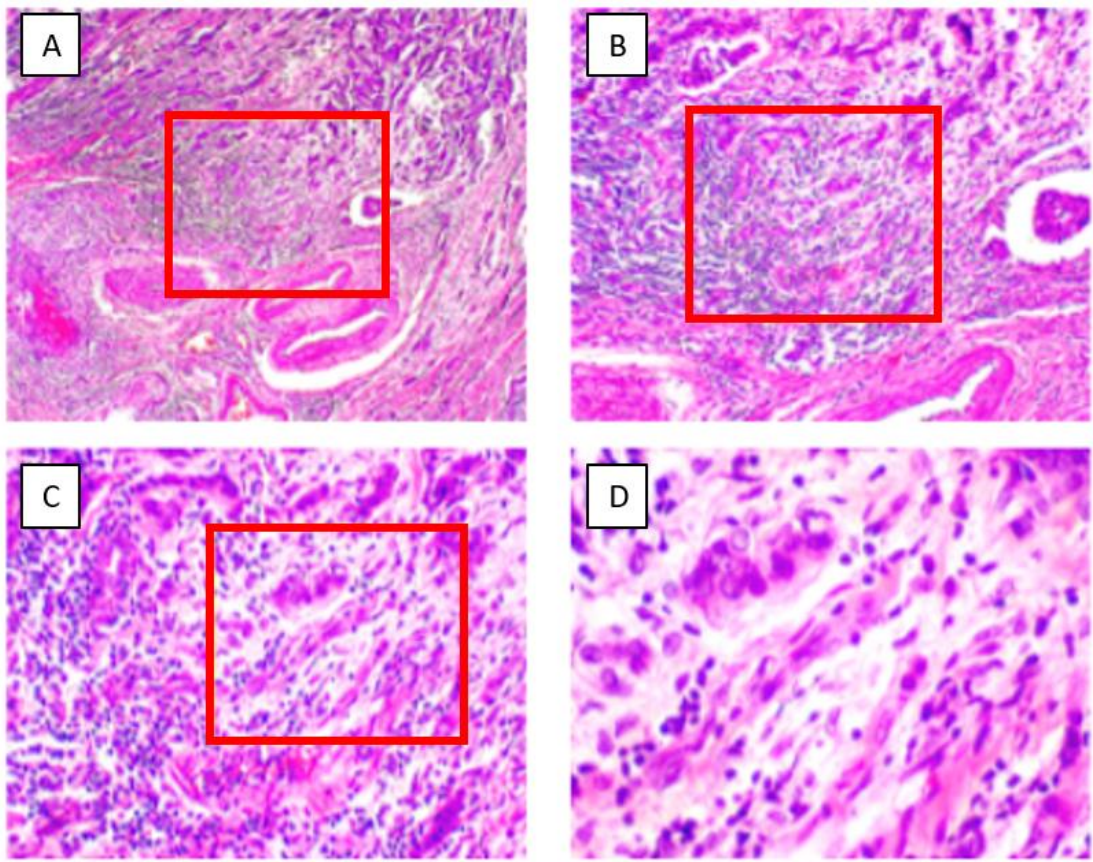

**Figure 2.** A slide of breast malignant tumor (stained with HE) extracted from the BreakHis dataset. Slides seen in different magnification factors: (A) 40X, (B) 100X, (C) 200X, and (D) 400X.

*2.2. Mammography*

Mammography is an X-ray based imaging modality and is the most commonly used radiographic imaging used by clinicians to screen for and detect BRCA [33 – 36], microcalcifications [37], and to perform image registration [38]. However, it has a significantly lower performance for BRCA detection in women with dense breast tissue, which have been shown to be at a higher risk of BRCA compared to women with non-dense breast tissue [39 - 42]. Image fusion of mammography images and other modalities, such as with Magnetic Resonance Imaging (MRI), can significantly improve diagnostic accuracy and lower abnormal tissue misclassification [33, 43]. Some of the most common mammography datasets are the Mammographic Image Analysis Society (MIAS) dataset, and the Curated Breast Imaging Subset of Digital Database for Screening Mammography (CBIS-DDSM). The MIAS dataset provides in depth information along with the diagnostic image, such as breast tissue type (fatty, fatty-glandular, dense-glandular), class of abnormality present (calcification, circumscribed mass, spiculated mass, Ill-defined mass, architectural distortion, asymmetry, or normal), tumor classification (benign, malignant), as well as X Y coordinates of center of mass and the mass radius. Figure 3 shows an arrangement of sample mammography images from the MIAS dataset.

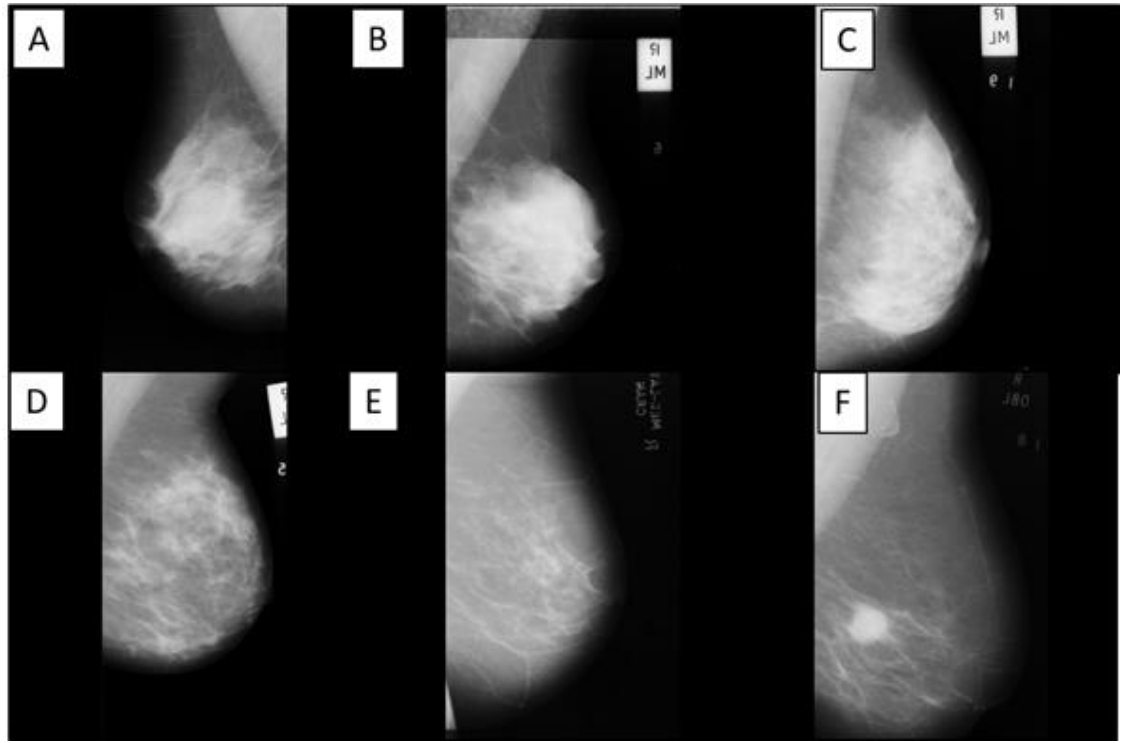

**Figure 3**. Sample of mammography images from MIAS dataset. (A) Fatty-glandular breast tissue presenting circumscribed benign mass [x:535, y:425, radius:197]. (B) Fatty-glandular breast tissue presenting circumscribed benign mass [x:522, y:280, radius:69]. (C) Dense breast tissue presenting no mass. (D) Fatty-glandular breast tissue presenting no mass. (E) Fatty-glandular breast tissue presenting no mass. (F) Fatty breast tissue presenting circumscribed malignant mass [x:338, y:314, radius:56].

*2.3.* Magnetic Resonance Imaging

Quantitative MRI is an essential imaging modality for cancer diagnosis in BRCA patients. Due MRI's flexible sensitivity and specificity to different types of BRCA, independent of breast density, it is an essential imaging modality when diagnosing BRCA in patients with genetic or familial tendencies to the disease [44].

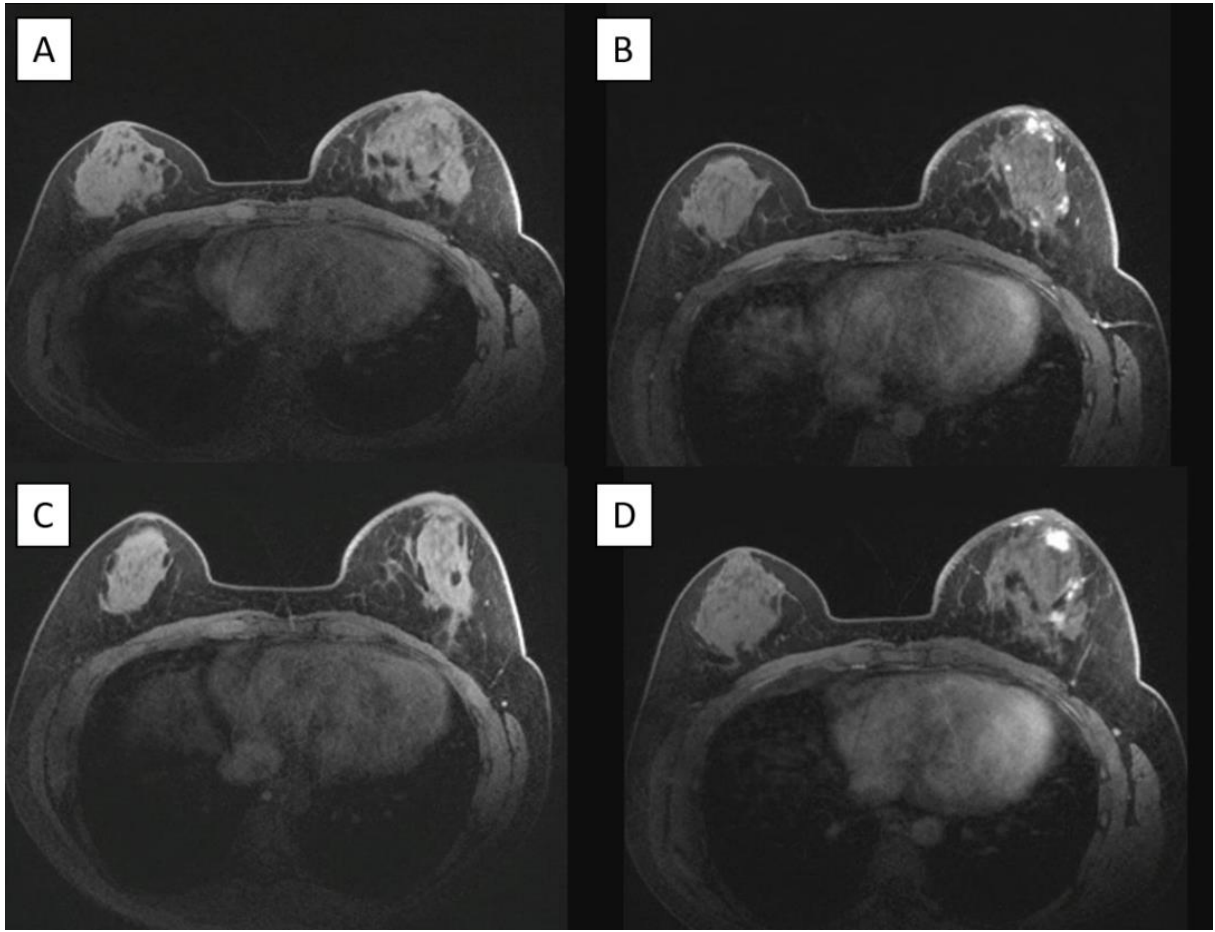

**Figure 4**. MRI images extracted from the QIN Breast DCE-MRI dataset. The dataset provides images from different patients under distinct conditions in the DICOM format, allowing us to use any of the 128 slices scanned from the examination.

*2.4. BRCA Prediction & Classification*

Many different characteristics commonly depicted in the literature [43] are considered when diagnosing and classifying BRCA and can vary from clinical risk factors to image-specific properties. Focusing on mammography and MRI modalities, the characteristics that are commonly used in the literature as features for the classification of BRCA are the following:

● Mass Radius - Computed through the average length of radial line segments from the center of the imaged mass to its border.

● Mass Perimeter – Computed as the linear distance around the mass border.

● Mass Area – Calculated by counting the number of pixels in the interior of the established mass border and summing half of the total number of pixels in the perimeter.

● Mass Compactness – Feature engineering from the perimeter and area, using the following formula:

$$C_{mass} = \frac{perimeter^2}{area}$$

● Mass smoothness – Quantified by calculating the difference between each individual radial length and the mean radial length.

● Mass Symmetry – Measured through the relative difference between line segments perpendicular to both sides of the major axis.

● Fractal Dimension – This feature can be approximated using the method proposed by *Mandelbrot et al.* which measures the mass border irregularity.

● Mass texture – Measured by evaluating the divergence of pixel tone concentrations inside the mass border.

**3. Radiomics**

Radiomics is a data-driven field within medical diagnostics and has more recently brought forth new potential solutions for accurate BRCA diagnosis, prognosis and treatment. With a combination of data mining from the patient file and image morphology from the diagnostic images, radiomics extracts extensible features providing quantifiable information that can be used for diagnosis [44 – 46]. Recent advances in the extraction of patient and disease information from diagnostics has permitted the field of radiomics to identify and define disease specific features for patient screening, diagnosis, prognosis, and treatment planning. Looking at the literature, we have seen radiomics pipelines applied to data mining, machine learning implementation, and image fusion.

*3.1.* Data mining

With the increase of digital imaging and diagnostic modalities in healthcare, an unprecedented amount of patient data has become available to clinicians and researchers alike. Extracting useful information from such large amounts of data has become a difficult task. Patient data, Imaging diagnostics, and features extracted from both can present structured and unstructured formats. Data analytics can be applied to diagnose, stage, and predict disease, as well as assist in decision making for clinicians [47]. The literature shows many possible approaches to this problem, and we have considered the following:

● C5.0: This algorithm makes use of several splitting algorithms which include entropy-based information gain. It works by splitting a sample into subsamples to obtain low-level samples. Low-level samples are selected and removed based on their contribution to the value. This algorithm is robust even when the dataset presents missing data or a large number of inputs and offers a simple yet powerful method of increasing the accuracy of classification.

● Fuzzy C-Means (FCM): This method allows a piece of data to belong to one of more clusters. Each data point is assigned to a cluster based on the Euclidean distance between the cluster center and the data point. The smaller the Euclidean distance between the data point and the cluster center, the higher its membership towards that particular cluster.

● K-means Clustering (KMC): A partitioning method, this algorithm is the best squared error based clustering algorithm. It works by assigning clusters and cluster centers correspondent to the centroid of the dataset. Objects are assigned clusters according to their similarity to other objects in the clusters.

● K-Nearest Neighbors (KNN): This instance based non-parametric learning algorithm memorizes the observations to classify the unseen test data. The algorithm compares the test observations with the nearest training observations.

● Naïve Bayes (NB): Considered one of the most effective statistical classification algorithms, it predicts class membership probabilities of each sample considering that the impact of an attribute value on a certain class does not depend of other attribute values.

● Partition Around Medoid (PaM): This algorithm is used to find medoids located in the center of clusters. Aimed at minimizing the average dissimilarity of objects to their closest selected object. In the same manner, we can minimize the dissimilarity between the object and the closest selected object. It works by first building an initial set of k objects, followed by exchanging selected objects with unselected objects to improve the quality of the clustering.

● Support Vector Machine (SVM): A machine learning technique based on statistical learning theory. This model generates a more complex boundary between classes, introducing margins on either side of the hyper plane. By maximizing margins, we can obtain the largest distance between the hyper plane and the samples.

*3.2.* Machine Learning

Recently, many machine learning approaches have been suggested to solve the ever growing data management problem within healthcare [48]. Amongst some of the most successful methods, we can note Recurrent Neural Networks (RNN) for biological signal processing, Convolutional Neural Networks (CNN) for Medical Image processing, Convolutional Long-Short-Term-Memory 2D (ConvLSTM2D) networks and Geometric Deep Learning (GDL) networks for medical video or image sequence processing, as well as many others.

*3.3.* Image Fusion

Medical Image Fusion is a process of combining two or more diagnostic images from the same or different modalities with the intention of improving the diagnostic quality of the image. Multi-modal medical image fusion encompasses several distinct techniques varying from image combinations to transformations. The process of medical image fusion can be broken down into two main stages, the first being image registrations, and the second being the fusion of relevant features from the registered images. For correctly performing medical image registration, it is important to consider spatial misalignment between different imaging modalities. This can be achieved through image rotation, scaling, and translation. In the case of BRCA diagnostic images, due to the soft tissue that compose the breasts, it is necessary to perform a nonrigid registration. Due to patient-specific breast morphology and heterogeneous nonlinear elastic properties of tissue in the breast, the implementation of deformable breast models is a current challenge [49]. A solution to this problem is the usage of fiducial skin markers along with a finite element method (FEM) software [50, 51].

The fusion of registered images requires accurate identification and selection of relevant features that are specific to the organ / disease under diagnosis. Presenting the information in a clinically relevant, user-friendly manner, as a single image is the main goal. However, some notable limitations are the dynamic range between images, and loss of information when fusing images from two distinct modalities. Most importantly, clinicians need to be able to easily and intuitively understand the resulting fused image and consider it an asset in the diagnostic process. Some current techniques include image color overlay [52, 53], intensity variation [54], and the usage of different colorspaces such as CIE XYZ, HSV, and HSL [55 – 56]. In the case of this project, Table 2 shows a summary of published medical imaging fusion methodologies that are relevant for BRCA diagnosis using mammography and MRI.

**Table 2**

Medical Image Fusion methods used with mammography and MRI, their applications, and algorithm strategy.

| Method | Application | Strategy |
| --- | --- | --- |
| Wavelets | Pseudo coloring, super resolution [57], medical diagnosis [58–61], feature level image fusion [62], segmentation [63], 3D conformal radiotherapy treatment planning [64] and color visualization [65] | Wavelet transform, multi resolution analysis [66], Discrete wavelet transform, Stationary wavelet transform, dual tree discrete wavelet transform |
| Knowledge | Segmentation [67], micro-calcification diagnosis [68], tissue classification [69], classifier fusion [70], breast cancer tumor detection [70,71], delineation & recognition of anatomical objects [67] and medical image retrieval [72–74] | Knowledge learning systems, expert systems |
| Fuzzy Logic | Cancer treatment [75], image segmentation and integration [75,76], maximization mutual information [77], tumor segmentation [78], image retrieval [79,80], spatial weighted entropy [79], feature fusion [79], multimodal image fusion [81 - 83], cancer diagnosis [84] | Neural networks, clustering neural network, fuzzy neural networks, wavelet neural networks |
| Neural Networks | Feature generation [85], classification [85], fusion [85,48,73,87,88,89–94], micro-calcification diagnosis [48], breast cancer detection [102,95,96], medical diagnosis [73,74,108], cancer diagnosis [97] | Image fuzzification, modification of membership values, Image defuzzification, fuzzy combination operators, neuro-fuzzy networks |

*3.4.* Clinical Decision Support

*3.4.1.* Diagnosis & Prognosis

Advanced diagnosis and prognosis of BRCA can be an effective way of reducing the patient's risk of life, as well as potentially increase patient longevity. Healthcare clinicians need rapid, accurate diagnosis to achieve these goals. Currently, many distinct approaches varying from classical statistical methods, image processing method, to machine learning methods are being used to identify and classify BRCA, as well as predicting the patients prognosis [98].

*3.4.2.* Surgical Planning

Since its inception, medical imaging diagnostics have assisted surgeons by increasing the safety and accuracy as well as decreasing the invasiveness of medical procedures. Image guided surgery systems can assist surgeons by performing data analysis for surgical planning [99 - 101] and surgical guidance [102 - 107]. In the case of BRCA CAD systems, information obtained through data can assist surgeons in surgical planning when preparing for a mastectomy or lumpectomy procedure in BRCA patients. Surgical planning systems present the surgeon with information gathered prior to the procedure, providing an array of information derived from fused patient data and imaging diagnostics. Some of the challenges of these systems are the registering of distinct image data, fusion of differently formatted data, and the selection of truly relevant information that can be considered an asset to the surgeon, as previously presented in section 2.3.

*3.4.3.* False-negatives

In Medical Imaging Diagnostics, the term false-negatives are exams which are incorrectly interpreted as negative, but the patient is diagnosed with BRCA within a certain timeframe (typically within 1 year), regardless of whether an abnormality is visible on the original examination or not [108]. Previous studies have shown the frequency of false-negative imaging diagnosis when correlated with a pathology examination to confirm the presence / lack of BRCA [108, 109]. These results are shown in Table 3 below.

**Table 3**

Frequency of True Positive (TP), False Positive (FP), True Negative (TN), and False Negative (FN), from diagnostic images of BRCA patients from sources [108, 109].

|  | (n=598) [109] | (n=23) [108] |
|---|---|---|
| True Positive | 39 | 6 |
| False Positive | 27 | 0 |
| True Negative | 511 | 5 |
| False Negative | 21 | 10 |

## 4. Image Segmentation

Image segmentation is a vital step in extracting relevant information from breast cancer images. In our proposed method, we considered special saliency in specific local regions, in order to extract information from the object of interest (tumor) and the surrounding tissue. The starting point is manually selected, and a conservative active contour algorithm delineates the tumor region. Next, the centroid and max distance of the borders are obtained in order to isolate a region of the image, which contains only tumor pixels. The border region, which contains both tumor pixels as well as surrounding tissue, is determined based on twice the maximum distance of the borders from the centroid. This region contains the gradient that will define the segmentation of tumor from surrounding tissue. Next, at the range of four times the maximum distance from the centroid, we select an area that contains only surrounding tissue. In order to quantify the pixel intensities for each selected area, and the contrast distribution of the background, we utilize a normalized histogram of the differences between the border region and the surrounding tissue region. Finally, we calculate the pixel saliency (for each specific pixel) in the border region through the following formula:

$$S(I_k) = S(c_l) = \sum_{j=1}^{n} f_j D(c_l, c_j),$$

where $c_l$ is the pixel value, $n$ is the number of possible shades of gray, and $f_j$ is the probability of the pixel with color $c_j$ being in the border region of the image. The distinct regions can be seen in Figure 5, below.

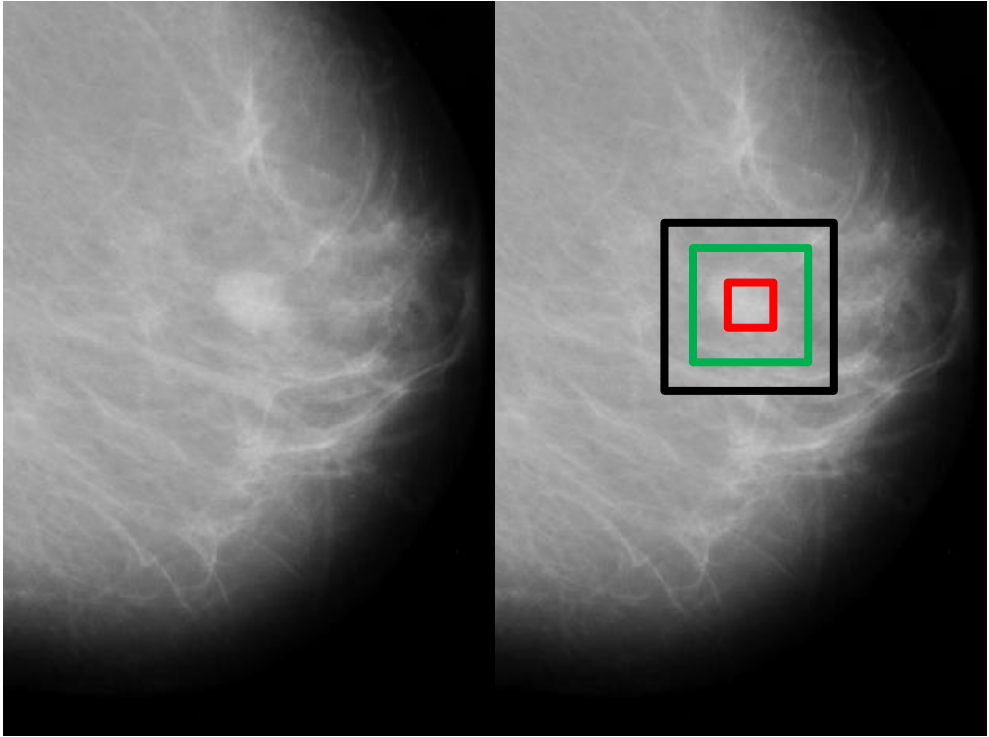

**Figure 5**. Distinct areas used during the feature extracting for the implementation of our proposed segmentation algorithm. Central Region (Red), Border Region (Green), Surrounding Region (Black).

We compared our developed method with two established segmentation algorithms, Active Contour (AC), and Region Growing (RG). The results can be seen below.

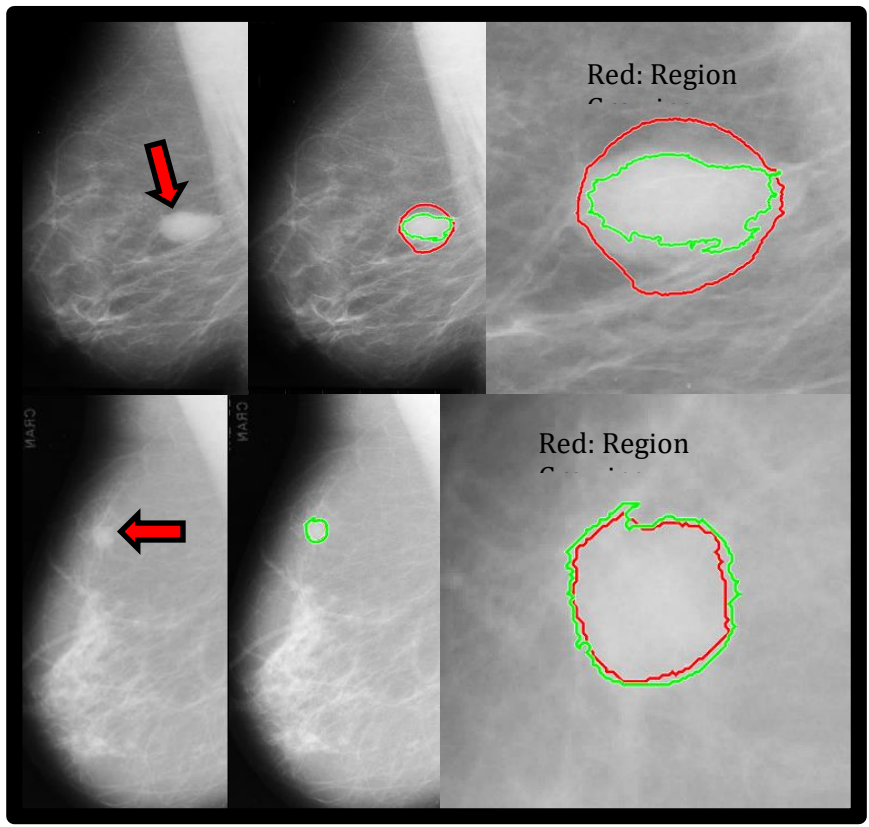

**Figure 6**. Comparison of our proposed segmentation algorithm with Region Growing method.

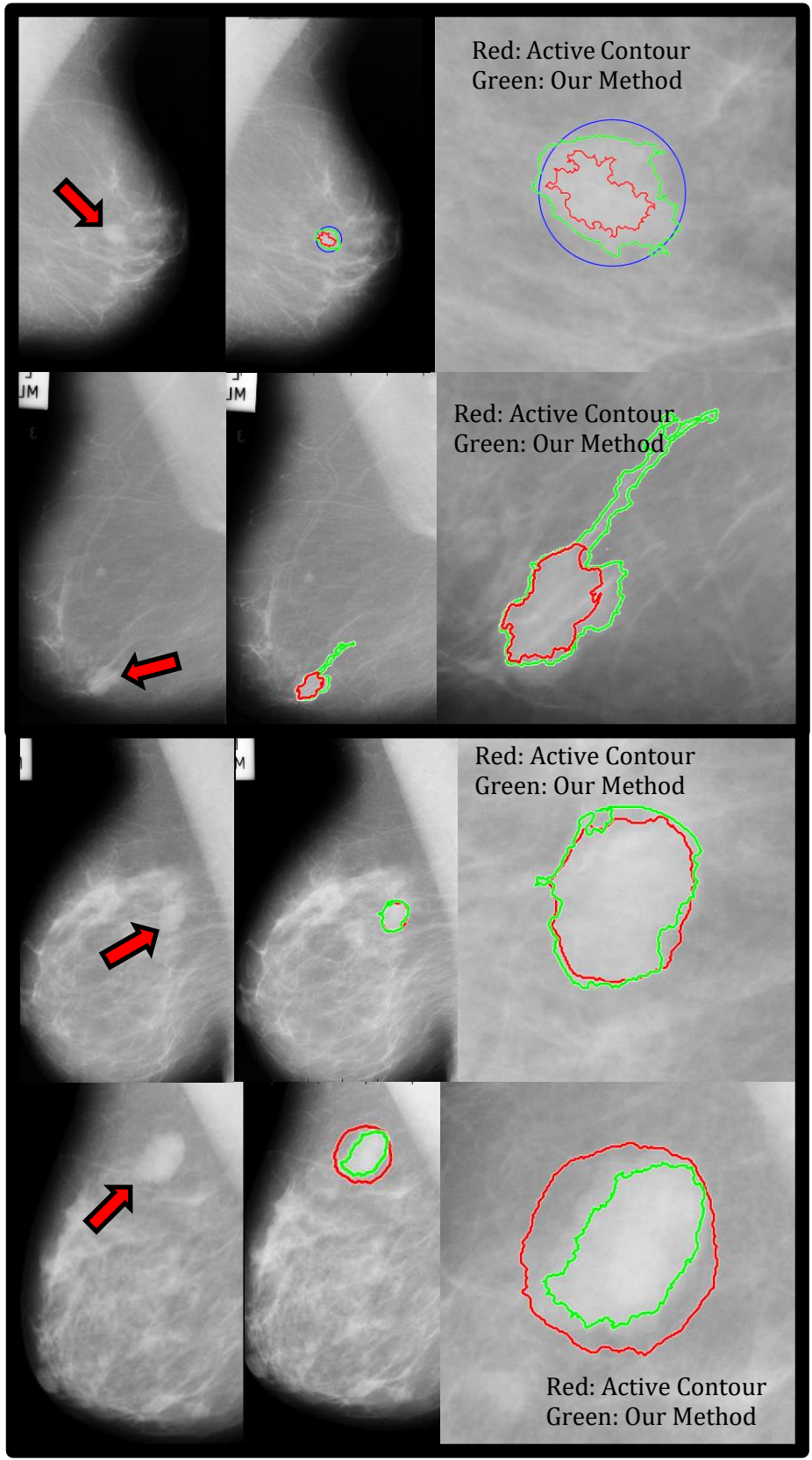

**Figure 7**. Comparison of our proposed segmentation algorithm with Active Contour method.

## 5. Conclusion

For well over two decades, imaging modalities have been used to assist doctors in the detection and diagnosis of, amongst other diseases, breast cancer. Some clear benefits of imaging diagnostics in breast cancer cases are non-invasive evaluation of internal masses, quantitative analysis of breast tissue and tumor masses, telemedicine, as well as digital breast imaging archives. In this paper, we present an overview of research in CAD methods for BRCA. We emphasized the distinction between the different imaging modalities involved in breast cancer diagnosis, and reviewed some of the most prominent feature extracting algorithms and machine learning methods in the present literature. In this review we also sought to provide a thorough and clear introduction to the multidisciplinary field of CAD systems, covering aspects from both radiology and computer vision. In particular, we classified recent publications related to computerized BRCA detection and diagnosis into distinct categories. In this report, we presented:

i) An extensive review of the literature involving breast cancer diagnosis, including imaging modalities such as histopathological images, MRI and mammography images.

ii) State of the art radiomic feature extraction approaches used to solve this problem with data structures and statistical analysis, such as C5.0, KNN, KMC, SVM, FCM, NB, and PaM,

iii) Some of the current machine learning and image processing methods, such as Wavelet Transform, Fuzzy Systems, Knowledge Learning Systems, Expert Systems, and Neural Networks.

iv) Finally, we proposed our breast cancer segmentation algorithm and presented some preliminary results. This segmentation algorithm will perform vital feature extraction from mammography images. The creation of a framework combining the aforementioned feature extraction, data-fusion, and machine learning methods capable of significantly reducing the false-negative rate of BRCA screening is of the utmost importance for future development of this field.